\documentclass[twocolumn,showpacs,preprintnumbers,amsmath,amssymb]{revtex4}

\usepackage{graphicx}
\usepackage{dcolumn}
\usepackage{bm}
\usepackage{epsfig}
\usepackage{color}

\begin{document}


\title{Structural relaxation and highly viscous flow}

\author{U. Buchenau}
 \email{buchenau-juelich@t-online.de}
\affiliation{%
Forschungszentrum J\"ulich GmbH, J\"ulich Centre for Neutron Science (JCNS-1) and Institute for Complex Systems (ICS-1),  52425 J\"ulich, GERMANY
}%

\date{February 1, 2018}

\begin{abstract}
The highly viscous flow is due to thermally activated Eshelby transitions which transform a region of the undercooled liquid to a different structure with a different elastic misfit to the viscoelastic surroundings. A self-consistent determination of the viscosity in this picture explains why the average structural relaxation time is a factor of eight longer than the Maxwell time. The physical reason for the short Maxwell time is the very large contribution of strongly strained inherent states to the fluidity (the inverse viscosity). At the Maxwell time, the viscous no-return processes coexist with the back-and-forth jumping retardation processes.  
\end{abstract}

\maketitle

\section{Introduction}

Within the last decades, the undercooled liquids close to the glass transition temperature have been the subject of ongoing discussions \cite{bnap,cavagna,bb}. The subject is still far from clear. 

In particular, it is not clear how much of the relaxation is due to viscous no-return processes and how much is due to back-and-forth jumps (retardation processes) \cite{visco}.

Some important facts are already well established: the dynamical heterogeneity \cite{hetero} (different regions of the undercooled liquid have different relaxation times); the existence of a terminal structural relaxation time \cite{greg,tinaa} after which the decay is exponential; the time-temperature superposition (one finds the same relaxation curves shifted to longer times at lower temperatures).

The time-temperature superposition works especially well in two vacuum pump oils, DC704 and PPE, which have developed into model glass formers because almost everything has been measured in them \cite{bo,tina}. DC704 consists essentially of four benzene rings loosely connected to a central silicon atom, PPE is a short chain of five phenylene rings.

In these two substances, the dielectric response, the adiabatic compressibility, the thermal expansion and the dynamic heat capacity move together with the shear response over a large temperature and relaxation time region \cite{bo}.

The shear response is characterized by the Maxwell time $\tau_M=\eta/G$ ($\eta$ viscosity, $G$ short time shear modulus) and by the peak frequency $\omega_G$ of the peak in $G''(\omega)$.

The condition $\omega_{peak}\tau=1$ defines a shear time  $\tau_G$, a dielectric $\tau_\epsilon$, an adiabatic compressibility relaxation time $\tau_\kappa$, a thermal expansion time $\tau_\alpha$ and a dynamic heat capacity relaxation time $\tau_{cp}$ (see Table I).

\begin{table}[htbp]
	\centering
		\begin{tabular}{|c|c|c|c|c|c|}
\hline
substance&$\tau_G/\tau_M$&$\tau_\kappa/\tau_M$&$\tau_\epsilon/\tau_M$&$\tau_\alpha/\tau_M$&$\tau_{cp}/\tau_M$\\
\hline   
DC704    & 0.78           & 3.39                &  3.83                 &        9.45         &          13.4     \\
PPE      & 0.83           & 3.64                &  5.63                 &                     &          12.3     \\
\hline
		\end{tabular}
	\caption{Sequence of relaxation times for different physical quantities in two vacuum pump oils \cite{bo}.}
	\label{tab:C}
\end{table}

Table I shows that the peak in $G''(\omega)$ is essentially determined by the Maxwell time.  The adiabatic compressibility relaxation time $\tau_\kappa$ is a factor of three to four longer than the Maxwell time. The terminal relaxation time, reflected in the dynamic heat capacity, is about a factor of thirteen longer than the Maxwell time. 

The aim of the present paper is to understand this factor of thirteen and to see which part of the relaxation is due to irreversible viscous processes.

The paper is based on the concept of thermally activated jumps between inherent states \cite{palmer,stillinger,scio,heuer} with a different structure. The shear response is attributed to the elastic shear misfit of the structures with respect to the viscoelastic surroundings.

The dominating role of the elastic shear misfit is suggested by the shear transformations of regions containing about twenty atoms, observed in the aging of a metallic glass \cite{atzmon1,atzmon2}.

After this introduction, Section II calculates the viscosity and the decay spectrum for these elastic shear misfit states. The results are compared to experimental data in Section III. Section IV discusses and summarizes the paper.

\section{Viscosity and viscous decay spectrum}

Consider a structural jump of a region over a barrier with energy $E_{B}$ into another structure with a different elastic misfit to the surroundings.

Before the jump, the region has a shear misfit angle $\epsilon_0$ (in radian) with respect to the surrounding viscoelastic matrix.  The region jumps into another shear misfit $\epsilon$. 

According to the Eshelby theory \cite{eshelby}, the shear energy increase or decrease by the jump is given by
\begin{equation}\label{delta0}
	\Delta=\frac{GNV\epsilon^2}{4}-\frac{GNV\epsilon_0^2}{4}.
\end{equation}
Here $NV$ is the volume of the region consisting of $N$ particles and $G$ is the short time shear modulus. Half of each of the two distortion energies is shear energy of the region, the other half is shear energy of the surroundings.

Let us define the shear states $\epsilon_0$ and $\epsilon$ by the dimensionless quantities $e_0$ and $e$ with
\begin{equation}
	e_0^2=\frac{GNV\epsilon_0^2}{4k_BT}\ \ \ \ e^2=\frac{GNV\epsilon^2}{4k_BT}.
\end{equation}

From the point of view of elasticity theory, the surroundings of the region react at short times like an isotropic elastic medium, describable by a strain tensor with one compression and five independent shear components. The shear misfits $e_0$ and $e$ are thus vectors in a five-dimensional shear misfit space.

In thermal equilibrium, the states $e$ in the five-dimensional shear misfit space have an average energy of 5/2 $k_BT$ in the normalized distribution
\begin{equation}
	p(e)=\frac{8}{3\sqrt{\pi}}e^4\exp(-e^2).
\end{equation}

The equation is based on the assumption \cite{ascom} of a constant density of stable structural states in distortion space and on the neglect of the difference in structural energy.

The underlying notion of more or less spherical rearranging regions is supported by recent nonlinear dielectric evidence \cite{chi5}.

The lifetime of the state $e_0$ is given by the rate with which it jumps into any other state (without return). Let us simplify the problem by assuming not only equal structural energies for all states, but also the same saddle point energy $E_B$ (without shear energy contribution) between all pairs of states.

$E_B$ must be rather large, of the order of $30k_BT$, to get the structural relaxation times up to the ms to s regime. Nevertheless, it will be seen in the following that the much smaller shear energy misfit contribution to the saddle point energy is of crucial importance.

The shear energy $E_s$ of the saddle point, supposed to lie in the middle between the two structures, is
\begin{equation}
	\frac{E_s}{k_BT}=\frac{1}{2}(e_0^2+e^2+2\vec{e_0}\cdot\vec{e}).
\end{equation}

The calculation of the escape rate from the state $e_0$ requires an integral over all possible $e$-values. In this integral, the contribution of the scalar product cancels. Therefore the effective barrier between $e_0$ and $e$ is changed in units of $k_BT$ by the amount $(e_0^2+e^2)/2-e_0^2=(e^2-e_0^2)/2$. 

Thus the jump rate from $e_0$ to $e$ gets a factor $\exp((e_0^2-e^2)/2)$ from the difference in the shear misfits.

With this, the state $e_0$ has the escape rate
\begin{align}\label{e0r}
	r=\frac{8}{3\sqrt{\pi}}r_0\int_0^\infty\exp((e_0^2-e^2)/2)e^4de \nonumber\\
	=4\sqrt{2}r_0\exp(e_0^2/2),
\end{align}
where the rate $r_0$ is given by
\begin{equation}\label{r0}
	r_0=\frac{N_s}{\tau_0\exp(E_B/k_BT)}.
\end{equation}
Here $\tau_0\approx 10^{-13}$ seconds and $N_s$ is the number of inherent states in the volume $8/3\sqrt{\pi}$ of
the five-dimensional $e$-space.

Note that the thermal occupation probability $p(e)$ does not enter eq. (\ref{e0r}), because the state $e_0$ is free to jump into any existing state.

Eq. (\ref{e0r}) shows that a strain-free region has a much longer lifetime than a strongly strained one, which can decay by fast jumps into lower-energy states.

The average lifetime in thermal equilibrium is obtained by the average escape rate
\begin{equation}
	r_0\int_0^\infty 4\sqrt{2}\exp(e^2/2)p(e)de=32r_0
\end{equation}
which yields the terminal relaxation time
\begin{equation}\label{tauc}
	\tau_c=\frac{1}{32r_0}.
\end{equation}

To calculate the viscosity, one needs the average squared shear angle of the jumps. Again, for a given $e_0$, this is
\begin{align}
	r_0\tau_c\frac{8}{3\sqrt{\pi}}\int_0^\infty(e_0^2+e^2)\exp((e_0^2-e^2)/2)e^4de \nonumber \\
	=\frac{\sqrt{2}}{8}\exp(e_0^2/2)(e_0^2+5).
\end{align}

Integrating over all values of $e_0$ with their thermal weight, one gets
\begin{equation}
	\frac{\sqrt{2}}{8}\int_0^\infty\exp(e^2/2)(e^2+5)p(e)de=10.
\end{equation}
This implies a rather large average squared shear angle per jump
\begin{equation}\label{eps2}
	\overline{\epsilon^2}=40\frac{k_BT}{GNV}.
\end{equation}

The calculation neglects the back-jump probability. But remember that the lifetime of the states is much longer than the shear stress relaxation time. The jumps are predominantly from one strongly strained state to another strongly strained state. Before the back-jump occurs, the strain of the state has dissolved.

In the Eshelby picture \cite{eshelby}, this implies that the average squared shear strain occurs not only in the inner volume $NV$, but also in an equivalent volume of the surroundings. Thus $\overline{\epsilon^2}$ is the average squared shear strain jump of the volume $2NV$.

The random walk of such jumps in the five-dimensional shear strain space leads for many jumps to the diffusion
\begin{equation}
	<\epsilon^2(t)>=10 Dt=\frac{\overline{\epsilon^2}t}{\tau_c}.
\end{equation}

The Einstein relation takes the form
\begin{equation}
	2NVD=\frac{k_BT}{\eta}
\end{equation}
which with eq. (\ref{eps2}) yields the final result
\begin{equation}\label{tcm}
	\frac{\tau_c}{\tau_M}=8.
\end{equation}

The result has been derived assuming the same barrier between all pairs of states. This is naturally not true. In reality, one must reckon with a continuous density of barriers, of which $E_B$ is merely the average value. This does not invalidate the calculation.

But $E_B$ is more than that: it is a crossover from flow to retardation processes. The integrated barrier density below $E_B$ does not suffice to start the flow. Therefore these lower barriers merely contribute back-and-forth jumps. 

The consideration implies $\tau_c\approx\tau_0\exp(E_B/k_BT)$ and this (with equs. (\ref{r0}) and (\ref{tauc})) implies $N_s\approx1/32$.

The value of $N_s$ determines the size of the regions. The connection is not straightforward, because $N_s$ is a density of possible structures in shear space, while the size of the region determines the number of possible structures of the region, without direct connection to their shape.

But, qualitatively, it is clear that the region size must increase with decreasing temperature, because the excess entropy of the undercooled liquid over the crystal decreases. With increasing size, $E_B$ will increase. This is one of the possible explanations for the fragility \cite{angell,bauer3}.

This explanation competes with the alternative one of a proportionality of $E_B$ to $G$ \cite{dyre}. We will come back to this dilemma \cite{bzr} in the discussion.

A nice feature of the calculation is that it allows to determine the viscous decay spectrum via equs. (\ref{e0r}) and (\ref{tauc}) for the lifetime of the state $e_0$ as a function of its shear strain energy. The relaxation time $\tau$ of state $e$ is given by
\begin{equation}\label{plt}
	\frac{4\sqrt{2}\tau_c}{\tau}=\exp(e^2/2).
\end{equation}

Translating $p(e)$ into a distribution $p(\ln{\tau})$ for the lifetimes in thermal equilibrium, one gets
\begin{equation}\label{pt}
	p(\ln{\tau})=\frac{\tau^2}{3\sqrt{2\pi}\tau_c^2}\left(\ln{\frac{4\sqrt{2}\tau_c}{\tau}}\right)^{3/2}.
\end{equation}

This viscous decay distribution function goes up to the value $4\sqrt{2}\tau_c$, the lifetime of a completely unstrained region. It is shown in Fig. 1.

\begin{figure}   
\hspace{-0cm} \vspace{0cm} \epsfig{file=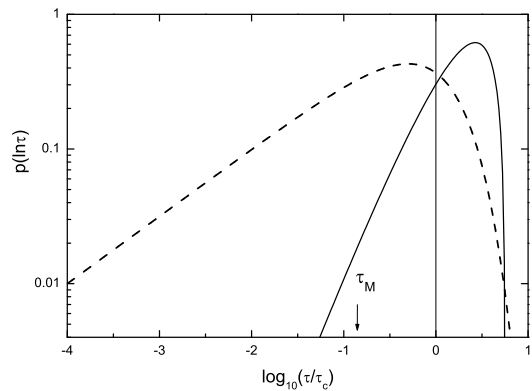,width=7 cm,angle=0} \vspace{0cm} \caption{The relaxation time distribution function $p(\ln{\tau})$ for the viscous structural decay of inherent states with different shear strains in thermal equilibrium (continuous line). The dashed line is the relaxation time distribution $p_{ret}(\ln{\tau})$ for the retardation processes (back-and-forth jumps).}
\end{figure}

Looking at Fig. 1, one gets an inkling where the factor thirteen comes from, because the peak centers around $1.6\tau_c=13\tau_M$.

Naturally, the viscous decay supplies only the irreversible part of the shear response, the viscous part which dominates at low frequencies. At higher frequencies, one has to reckon with back-and-forth jumps, which provide the reversible part of the shear response and give rise to the recoverable compliance \cite{visco,ascom}.

We do not yet have a theoretical calculation for the reversible part of the spectrum, but one can make use of the pragmatical solution of Schr\"oter and Donth \cite{donth} for glycerol: the retardation processes can be described in terms of the Cole-Davidson function with $\beta\approx1/2$ and $\tau_{CD}\approx\tau_c=8\tau_M$, which has the negative logarithmic slope -1/2 at high frequency and a Debye-like cutoff at $\tau_c$. 

For our purpose, it is more appropriate to express the Cole-Davidson behavior in terms of a relaxation time retardation function
\begin{equation}\label{ret}
	p_{ret}(\ln{\tau})=p_0\left(\frac{\tau}{\tau_c}\right)^\beta\exp(-\tau/\tau_c)
\end{equation}
(see Fig. 1). One already knows the average lifetime $\tau_c$ from the viscosity consideration. This leaves only the Kohlrausch parameter $\beta$ and $p_0$ as fitting parameters.

For a Kohlrausch $\beta=1/2$, $p_0$ determines the recoverable compliance $J_0$ via
\begin{equation}
	GJ_0-1=p_0\frac{\sqrt{\pi}}{2}.
\end{equation}

Though one still works on a semi-phenomenological level, the theoretical consideration presented here has reduced the number of fitting parameters for $G(\omega)$ to four: the short time shear modulus $G$, the viscosity $\eta$, the Kohlrausch $\beta$, and the recoverable compliance $J_0$.

One gains another advantage: one can fit the relaxation function of other physical variables with the spectra $p(\ln{\tau})$ and $p_{ret}(\ln{\tau})$ obtained from $G(\omega)$.

To do this, one defines a combined spectrum
\begin{equation}\label{comb}
	p_{comb}(\ln{\tau})=p(\ln{\tau})+w_rp_{ret}(\ln{\tau})/p_0,
\end{equation}
where the parameter $w_r$ characterizes the relative weight of the retardation spectrum.

As Table I shows, other physical quantities can equilibrate much earlier than the terminal relaxation. This can be described by a (physical-quantity dependent) maximal relaxation time $\tau_{max}$.

One then calculates the relaxation function of the quantity by the relaxation time distribution $p_{comb}(\ln{\tau})$ between $\tau=0$ and $\tau_{max}$.

This program will be carried out for PPE in the next section. 

\section{Comparison to experiment}

Fig. 2 shows the fit of the $G(\omega)$-data \cite{bo,tina} of PPE at 252.5 K. The parameters are given in the caption of Fig. 2.

\begin{figure}   
\hspace{-0cm} \vspace{0cm} \epsfig{file=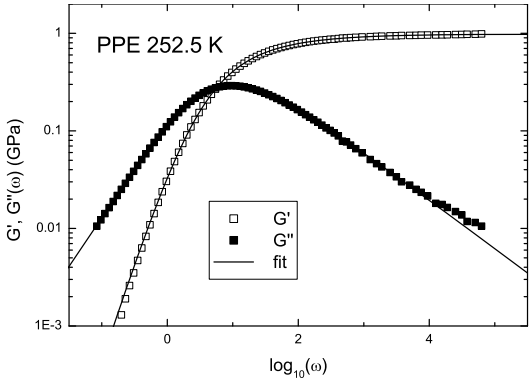,width=7 cm,angle=0} \vspace{0cm} \caption{Fit of $G(\omega)$-data \cite{bo,tina} of PPE at 252.5 K with $G=0.99\ GPa$, $\eta=0.129\ GPas$, $\beta=1/2$ and $GJ_0=2.65$.}
\end{figure}

Having $G$, $\eta$, $\beta$ and $J_0$, the calculation is done in the following way: one has $\tau_c=8\tau_M=8\eta/G$. The prefactor $p_0$ for $p_{ret}(\ln{\tau})$ is calculated from
\begin{equation}
	(GJ_0-1)=\int_{-\infty}^\infty p_0\left(\frac{\tau}{\tau_c}\right)^\beta\exp(-\tau/\tau_c)d\ln{\tau}.
\end{equation}

For a given $\omega$, one has the shear compliance
\begin{equation}
	GJ(\omega)=\int_{-\infty}^\infty \frac{p_{ret}(\ln{\tau})}{1+i\omega\tau}d\ln{\tau}-\frac{1}{i\omega\tau_M}
\end{equation}
and obtains $G(\omega)=1/J(\omega)$.

\begin{figure}   
\hspace{-0cm} \vspace{0cm} \epsfig{file=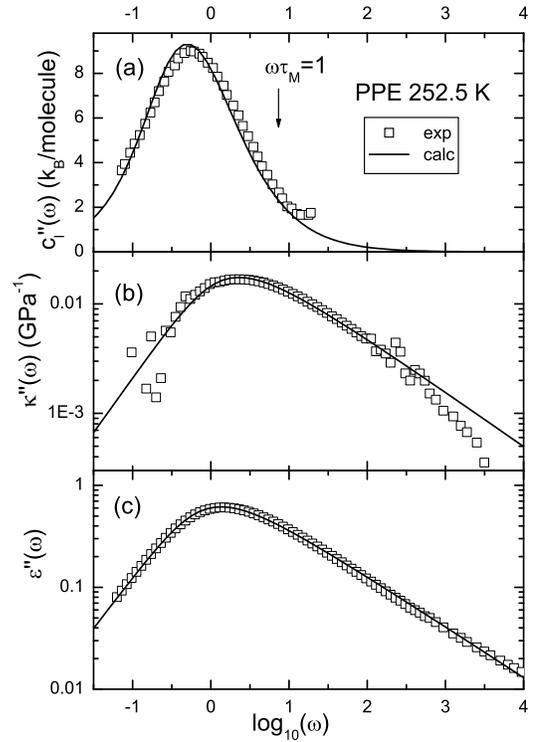,width=7 cm,angle=0} \vspace{0cm} \caption{Comparison of the combined relaxation time distribution of eq. (\ref{comb}) to data \cite{bo,tina,cp} measured at 252.5 K in PPE a) dynamic heat capacity with $\Delta c_l=22.5\ k_B/molecule$, $w_r=0$ and no upper limitation of the spectrum b) dynamic compressibility with $\Delta\kappa=0.053\ GPa^{-1}$, $w_r=0.3$ and $\tau_{max}=\tau_c$ c) dielectric relaxation with $\Delta\epsilon=1.86$, $w_r=0.32$ and $\tau_{max}=1.73\tau_c$.}
\end{figure}

Fig. 3 shows the fits of other physical quantities in terms of the combination of the two $G(\omega)$-spectra of eq. (\ref{comb}). The parameters are again listed in the caption.

The exact procedure is exemplified for the dielectric case. One has $\tau_c$ from the $G(\omega)$-data. One further needs $\Delta\epsilon=\epsilon_s-\epsilon_\infty$ (the difference between the static dielectric susceptibility $\epsilon_s$ and the high frequency limit $\epsilon_\infty$), the retardation weight factor $w_r$ and the maximum relaxation time $\tau_{max}$.

Then one can calculate $p_{comb}(\ln{\tau})$ from eq. (\ref{comb}) and normalizes it to 1 with the condition
\begin{equation}
	\int_{-\infty}^{\ln{\tau_{max}}}p_{comb}(\ln{\tau})d\ln{\tau}=1.
\end{equation}

$\epsilon(\omega)$ is given by
\begin{equation}
\epsilon(\omega)=\epsilon_\infty+\Delta\epsilon\int_{-\infty}^{\ln{\tau_{max}}}\frac{p_{comb}(\ln{\tau})}{1+i\omega\tau}d\ln{\tau}.
\end{equation}

This procedure turns out to be easiest for the dynamic heat capacity of PPE \cite{bo,cp} in Fig. 3 (a), with $w_r=0$ and no upper lifetime cutoff, with only $\Delta c_p$ as adaptable parameter. This provides a very good description of the measured data.

The reason why one does not see the retardation processes in the dynamic heat capacity is revealed in the work of Christensen et al \cite{tage}. 

The measurement is hampered by its mechanical conditions; one does not measure the real dynamic heat capacity $c_p(\omega)$, but rather a longitudinal dynamic heat capacity $c_l(\omega)$. At the low frequency end, $c_l(\omega)$ equals $c_p(\omega)$, but at the high frequency end $c_l''(\omega)$ should be lower by a factor of about 2/3 (estimated for glycerol \cite{tage}).

From these results, it seems almost certain that one would see a retardation component if one could measure $c_p(\omega)$ itself.

But, even so, the fact that $c_l(\omega)$ does indeed reflect the full $p(\ln{\tau})$ of eq. (\ref{pt}) up to its upper cutoff provides a strong support for the calculations of Section II. In units of the independently measured $\tau_M$, one finds the peak exactly at the place where it should be, with the correct width.

The adiabatic compressibility data in Fig. 3 (b) and the dielectric data in Fig. 3 (c) require the $w_r$ and $\tau_{max}$ values given in the caption of Fig. 3. Note that the two $w_r$-values agree within the error limit.

\section{Discussion and Summary}

The paper presents a new view of the structural relaxation in undercooled liquids, attributing the viscous flow to Eshelby transitions between inherent states with different elastic shear misfit to the surrounding viscoelastic matrix.

Since the explanation limits itself to irreversible structural transitions, it is not able to reproduce the back-and-forth jumps occurring at times longer than the structural lifetime. These are put in by hand, defining them by a Kohlrausch $\beta$ close to 1/2 and a recoverable compliance $J_0$.

The explanation reverses the traditional Debye-Stokes-Einstein one \cite{niss1}, which considers the shear stress relaxation with the subsequent viscous flow as the cause and the rotation of molecules in the viscous liquid as the consequence.

In the present explanation, the structural decay of a whole region, involving a rotation of many molecules together, is the cause. The shear fluctuations accompanying this cause are so strong that one gets a small Maxwell time as a consequence. 

But there is no Debye-Stokes-Einstein mechanism, because there is no viscous flow in the small volume of a single molecule. The mechanism might work at higher temperatures \cite{schober}, but not close to the glass transition.

A similar consideration for the local shear modulus offers a way out of the fragility dilemma \cite{angell,dyre,bzr}: the macroscopic shear modulus $G$ owes its strong temperature dependence \cite{dyre} to local shear softening. These soft spots sit in the strongly distorted regions and in the grain boundaries, not in the stable unstrained regions responsible for $\tau_c$. Therefore the fragility needs not be due to the macroscopic $G$.

In this picture, nothing special happens at the Maxwell time, where one only has a mixture of the decay of the strongly shear strained regions with back-and-forth jumps.

The scheme allows to calculate a lifetime distribution for the states, with no other input as the Maxwell time.

But, as usual in the field, this partial answer does not solve the full riddle. The calculation of the retardation component is missing.

Though it seems obvious that flow and retardation are two sides of the same medal, this task is not trivial. One cannot simply assume a constant barrier density toward lower barriers, because the retardation barrier density decreases according to the Kohlrausch law. There must be a physical reason for this universal decrease, but a quantitative analysis is missing.

The possibility to describe other physical quantities with a combined decay-retardation spectrum and an upper cutoff time shows that some of them are able to equilibrate before the last region decays. Though there are still fluctuations, there is no longer a dissipation. One concludes that the field within the sample for this physical quantity must be already zero. 

To summarize, the present treatment of undercooled liquids focuses on the shear misfit of local inherent states as the decisive variable. The approach is able to explain why the shear stress relaxation precedes the structural relaxation by a decade and opens a new way to describe measured relaxation data.


\begin{thebibliography}{99}
\bibitem{bnap} R. B\"ohmer, K. L. Ngai, C. A. Angell, and D. J. Plazek, J. Chem. Phys. {\bf 99}, 4201 (1993)
\bibitem{cavagna} A. Cavagna, Phys. Rep. {\bf 476}, 51 (2009)
\bibitem{bb} L. Berthier and G. Biroli, Rev. Mod. Phys. {\bf 83}, 587 (2011) 
\bibitem{visco} U. Buchenau, Phys. Rev. E {\bf 95}, 062603 (2017)
\bibitem{hetero} R. Richert, J. Phys.: Condens. Matter {\bf 14}, R703 (2002)
\bibitem{greg} G. B. McKenna, J. Non-Cryst. Solids {\bf 172-174}, 756 (1994)
\bibitem{tinaa} T. Hecksher, N. B. Olsen, K. Niss, and J. C. Dyre, J. Chem. Phys. {\bf 133}, 174514 (2010)
\bibitem{bo} B. Jakobsen, T. Hecksher, T. Christensen, N. B. Olsen, J. C. Dyre, and K. Niss, J. Chem. Phys. {\bf 136}, 081102 (2012)
\bibitem{tina} T. Hecksher, N. B. Olsen, K. A. Nelson, J. C. Dyre and T. Christensen, J. Chem. Phys. {\bf 138}, 12A543 (2013)
\bibitem{palmer} R. G. Palmer, Adv. Phys. {\bf 31}, 669 (1982)
\bibitem{stillinger} P. G. Debenedetti and F. H. Stillinger, Nature {\bf 410}, 259 (2001)
\bibitem{scio} S. Mossa, E. La Nave, F. Sciortino, and P. Tartaglia, Eur. Phys. J. B {\bf 30}, 351 (2002)
\bibitem{heuer} A. Heuer, J. Phys.: Condens. Matter {\bf 20}, 373101 (2008)
\bibitem{atzmon1} J. D. Ju, A. Nwankpa, and M. Atzmon, J. Appl. Phys. {\bf 109}, 053522 (2011)
\bibitem{atzmon2} J. D. Ju and M. Atzmon, MRS Communications {\bf 4}, 63 (2014)
\bibitem{eshelby} J. D. Eshelby, Proc. Roy. Soc. {\bf A241}, 376 (1957)
\bibitem{ascom} U. Buchenau, J. Chem. Phys. {\bf 134}, 224501 (2011)
\bibitem{chi5} S. Albert, Th. Bauer, M. Michl, G. Biroli, J.-P. Bouchaud, A. Loidl, P. Lunkenheimer, R. Tourbot, C. Wiertel-Gasquet, and F. Ladieu, Science {\bf 352}, 1308 (2016)
\bibitem{angell} C. A. Angell, J. Res. Natl. Inst. Stand. Technol. {\bf 102}, 171 (1997)
\bibitem{bauer3} Th. Bauer, P. Lunkenheimer and A. Loidl, Phys. Rev. Lett. {\bf 111}, 225702 (2013)
\bibitem{dyre} J. C. Dyre, Rev. Mod. Phys. {\bf 78}, 953 (2006)
\bibitem{bzr} U. Buchenau, R. Zorn, and M. A. Ramos, Phys. Rev. E {\bf 90}, 042312 (2014)
\bibitem{donth} K. Schr\"oter and E. Donth, J. Chem. Phys. {\bf 113}, 9101 (2000) 
\bibitem{cp} B. Jakobsen, N. B. Olsen, and T. Christensen, Phys. Rev. E {\bf 81}, 065505 (2010)
\bibitem{tage} T. Christensen, N. B. Olsen, and J. C. Dyre, Phys. Rev. E {\bf 75}, 041502 (2007)
\bibitem{niss1} B. Jakobsen, K. Niss, and N. B. Olsen, J. Chem. Phys. {\bf 123}, 234510 (2005)
\bibitem{schober} H. R. Schober and H. L. Peng, Phys. Rev. E {\bf 93}, 052607 (2016)
\end{thebibliography}
\end{document}